\documentclass[aps,pra,twocolumn,showpacs,superscriptaddress,groupedaddress]{revtex4}
\usepackage[colorlinks=true, linkcolor=red, citecolor=blue, urlcolor=blue]{hyperref}
\usepackage[utf8]{inputenc}
\usepackage{graphicx}
\usepackage{amssymb}
\usepackage{amsmath}
\usepackage{dsfont}

\newcommand{\ex}[1]{{\langle{#1}\rangle}}

\begin{document}
\title{Tests against noncontextual models with measurement disturbances}

\author{Jochen Szangolies}
\email{szangolies@thphy.uni-duesseldorf.de}
\affiliation{Institut f\"ur Theoretische Physik III, 
Heinrich-Heine-Universit\"at D\"usseldorf, D-40225 D\"usseldorf, Germany}
\affiliation{Naturwissenschaftlich-Technische Fakult\"at, Universit\"at Siegen, Walter-Flex-Straße 3, D-57068 Siegen, Germany}
\author{Matthias Kleinmann}
\email{matthias.kleinmann@uni-siegen.de}
\author{Otfried G\"uhne}
\email{otfried.guehne@uni-siegen.de}
\affiliation{Naturwissenschaftlich-Technische Fakult\"at, Universit\"at Siegen, Walter-Flex-Straße 3, D-57068 Siegen, Germany}

\begin{abstract}
The testability of the Kochen-Specker theorem is a subject of ongoing 
controversy. A central issue is that experimental implementations relying on 
sequential measurements cannot achieve perfect compatibility between the
measurements and that therefore the notion of noncontextuality does not apply. 
We demonstrate by an explicit model that such compatibility violations 
may yield a violation of noncontextuality inequalities, even if we assume that 
the incompatibilities merely originate from context-independent noise.  We 
show, however, that this problem can be circumvented by combining the ideas 
behind Leggett-Garg inequalities with those of the Kochen-Specker theorem.
\end{abstract}
\maketitle

\section{Introduction} 

Bell's theorem \cite{Bell} is a famous no-go result that provides  
constraints on the program of interpreting quantum mechanics as 
an incomplete theory in the sense of Einstein, Podolsky, and 
Rosen \cite{EPR}. It is expressed via inequalities that are 
fulfilled by any local realistic theory, but which are predicted 
to be violated by quantum mechanics. Experimentally, it is 
indeed found that quantum mechanics violates these inequalities 
for certain entangled states \cite{Asp,Zei}. Similar to Bell, Leggett and Garg 
\cite{LG} have proposed inequalities that are fulfilled by theories that 
satisfy a criterion of \emph{macroscopic realism}, meaning that a system always 
occupies one of the states accessible to it. Under the further assumption of 
measurement non-invasiveness, the correlations between measurements performed 
on the system at different points in time obey a bound that is violated by 
quantum mechanics.

A third no-go result is provided by the Kochen-Specker theorem 
\cite{KS, BKS}. Essentially, it replaces Bell's assumption of locality 
with the condition of noncontextuality: the outcome of a measurement 
on a system should not depend on other compatible measurements performed 
on the same system. Here, two measurements are called compatible, if they
can be measured simultaneously or in a temporal sequence without any disturbance. 
The Kochen Specker result contains Bell's theorem as a special case in 
which the measurements are performed at spatial separation \cite{Mermin}. 
It is, however, also applicable to single quantum systems; consequently, 
entanglement is not necessary for violations of noncontextuality. In fact, 
violations of Kochen-Specker inequalities occur for all quantum systems of 
dimension $d\geq3$, independent of the initial quantum state \cite{KS}. 

However, in contrast to Bell's theorem, the Kochen-Specker theorem does not 
readily lend itself to experimental tests of quantum mechanics. The direct 
testability is stymied by the fact that, during each experiment, only one 
measurement context is accessible at any given time. This limitation can be 
overcome, however, because similarly to Bell's theorem, the Kochen-Specker 
theorem can be expressed using inequalities, though it was originally not cast 
into this form \cite{Klyachko, Adan}. This permits testing quantum 
contextuality by using measurements that are carried out sequentially. 

But even in this formulation, the question of experimental testability poses 
further difficulties. The reason for this lies in the notion of contextuality, 
which only applies in the case of compatible observables.  But in an experiment, this 
condition cannot be fulfilled perfectly; indeed, even measuring the same observable twice 
may yield different results.  This is due to the 
unavoidable presence of noise during the measurement process, which leads to 
disturbances of the state. It has been argued that this inherent difficulty, 
together with a related issue concerning the finite precision of real 
measurements, nullifies the physical significance of the Kochen-Specker theorem 
\cite{errterms, fin1, fin2}.

Different strategies have been proposed to overcome this problem.  In 
Ref.~\cite{errterms}, the  modification of Kochen-Specker inequalities through 
the introduction of error terms was considered (see also Ref.~\cite{nature}). 
Given that the measurement-induced disturbances are cumulative, these terms 
compensate for the violations of compatibility. A related approach,  addressing 
a similar loophole in experimental tests of Leggett-Garg inequalities, was 
proposed in Ref.~\cite{clums}. On the other hand it was suggested in 
Ref.~\cite{qutrits} to perform experiments on separate qutrits in order to 
forestall the possibility of violations of compatibility. 

In this paper, we take a different approach. First, we consider the question: what does an experimentally observed violation of a noncontextuality inequality license us to conclude? By proposing an explicit model capturing the effects of noise-induced compatibility violations, we argue that 
to conclude contextuality from the violation alone is difficult to justify: the model produces violations of noncontextuality 
inequalities while being independent of the measurement context, 
and thus, noncontextual in this sense. In particular, we show 
that even the introduction of error terms as proposed in Ref.~\cite{errterms} cannot settle the issue.

We then propose a way to circumvent this problem by taking into 
account the ideas of Leggett and Garg: imposing a suitable time-ordering onto the measurements, it turns out to be possible to formulate inequalities that cannot be violated within our 
framework, and thus, allow to rule out more general hidden-variable models
under realistic experimental conditions.

This paper is organized as follows. In Section II, we briefly recall the notions of compatibility and contextuality. Then, we propose an explicit classical noise model capable of inducing compatibility violation in such a way as to violate contextuality inequalities. In Section III we 
show that the model cannot be ruled out by previous approaches. To remedy this problem, in Section IV we propose new inequalities, utilizing ideas from Leggett and Garg. These inequalities allow to rule out more general hidden variable models.

\section{Noncontextual models}
As a basis for our investigations we take a variant of the well-known 
Clauser-Horne-Shimony-Holt (CHSH) inequality \cite{CHSH}
\begin{equation}\label{CHSH}
 \ex{\chi_{\mathrm{CHSH}}}=\ex{AB}+\ex{BC}+\ex{CD}-\ex{DA}\stackrel{\textit{NCHV}}{\leq}2\stackrel{\textit{QM}}{\leq}2\sqrt{2}.
\end{equation}
Each of the observables $A$, $B$, $C$, $D$ has outcomes $\pm1$, and $\ex{AB}$ 
denotes the average over many repetitions obtained by measuring first $A$,
then $B$, and then multiplying the results.
If we assume that the observables in each context $\ex{AB}$, etc., are 
compatible, then \textit{NCHV} denotes the classical (noncontextual 
hidden-variable) bound, i.e., the value obtained if each of the observables is 
assumed to have a fixed value independently of which context it is measured 
in.  The bound \textit{QM} denotes the maximal value achievable in quantum 
mechanics \cite{TsB}.
The question now is: suppose one experimentally observes $\ex{\chi_{\mathrm{CHSH}}}>2$. Is this sufficient to conclude contextual behavior?

First, we need to make the notions of compatibility and noncontextuality 
precise. Consider some observables $\mathcal{O}=\lbrace A, B, C, 
\ldots\rbrace$. Compatibility then means that within any sequence of 
measurements composed of these observables, the observed value does not depend 
on the point at which it is measured within the sequence. That is, for any 
sequence of compatible measurements $\mathcal C$, the observed value of $O$ at 
the $i$th point in the sequence, $v_i(O|\mathcal{C})$, does not depend on $i$, 
i.e., $v_i(O|\mathcal{C}) = v_j(O|\mathcal{C})$ for all $i$ and $j$. This 
formalizes the notion that measurement of one observable does not influence the 
measurement of any other observable.
 
Then, any set of compatible observables $\mathcal C$ is called a 
\emph{context}.  A theory is called \emph{noncontextual}, if for all 
observables $O$ and for all contexts $\mathcal{C},\mathcal{C}^\prime$ the 
observed value is independent of the context, i.e., 
$v(O|\mathcal{C})=v(O|\mathcal{C}^\prime)$.
Note that through the definition of a context, the notion of noncontextuality explicitly depends on compatibility.

To approach the question, we construct a counterexample given by a simple model for noise-induced disturbances of the hidden-variable states. 
These hidden states $\lambda_i$ are assumed to completely specify all possible experimental outcomes. In the present case, they can thus be indexed 
by the dichotomic outcomes of measurements of the observables $\mathcal{O}=\lbrace A, B, C, D\rbrace$: a given state is specified uniquely by a 
set of values $v(O)\in\lbrace\pm1\rbrace$ for all $O \in\mathcal{O}$. For ease 
of notation, this set of values may be interpreted as a binary number, whose 
decimal value is used to index the state, i.e., $\lambda_2=\lambda_{(++-+)}$ 
denotes the state that produces the measurement results $A=+1$, $B=+1$, $C=-1$ 
and $D=+1$.  The model can be generalized by considering states that are convex 
combinations of the value attributions $\lambda_i$, such that the most general 
state can be written as a mixture $\sum_{i=0}^{(2^n)-1}p_i\lambda_i$, where 
$\sum_{i=0}^{(2^n)-1}p_i=1$ and $n$ denotes the number of observables. 

The dynamics of this model now is such that after every measurement, the system may randomly execute a transition to a different state. 
Note that this transition does not depend on which measurement was carried out. 
This models the effect of noise introduced during measurement, i.e., after a 
noisy interaction with the system, further measurements will in general yield 
different results. We will now show that this is equivalent to the introduction 
of compatibility violations in a realistic experiment, and, crucially, that 
these violations may lead to false positives in Kochen-Specker tests.

\begin{figure}
\centering
\includegraphics[width=\columnwidth]{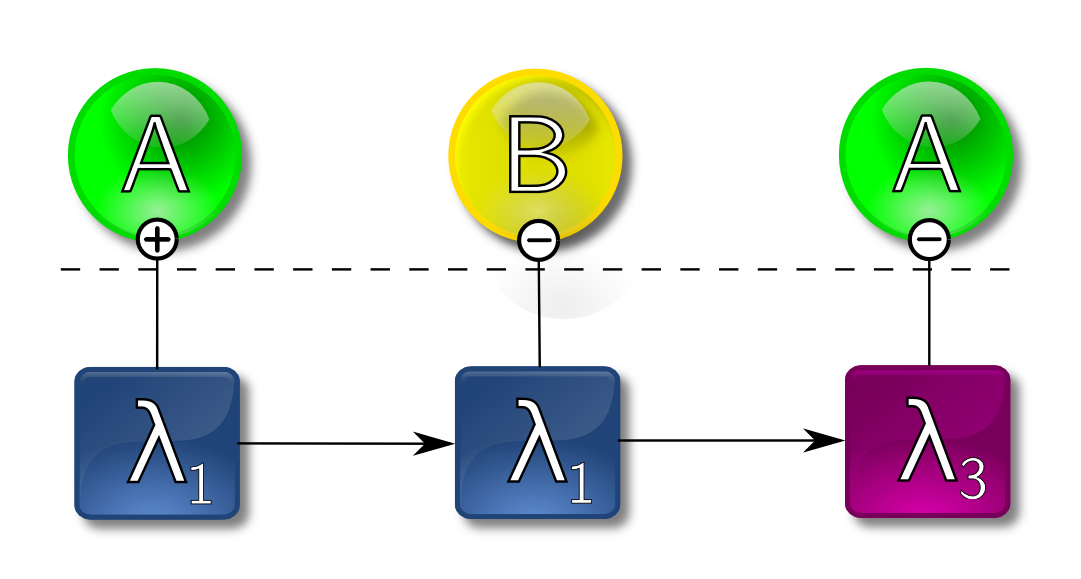}
\caption{Schematic representation of a sequence of measurements. Measurements $A$ and $B$ are performed sequentially on a system whose 
(hidden-variable) state $\lambda$ evolves stochastically as indicated. Time runs left to right.}
\label{KD1}
\end{figure}

Consider the evolution depicted in Fig.~\ref{KD1}: a measurement of the 
observable $A$ is made on a system in the state $\lambda_1$, consequently 
producing the result $+1$. Subsequently, the observable $B$ is measured, 
yielding $-1$. Then, the system undergoes a state transition to $\lambda_3$, 
and a subsequent measurement of $A$ yields $-1$. Thus, compatibility is 
violated.

Of course, this model cannot suffice to capture all quantum mechanical effects; in particular, for \emph{a priori} incompatible observables, it is 
easy to show that its behavior differs from that of quantum mechanics: take a measurement sequence such as $AAA$. Without disturbances, both quantum mechanics and the model 
predict that the same result will be repeated three times; allowing for noise influences, there will be a small probability of disagreement. 
Measuring $ACA$, however, since $A$ and $C$ are not compatible, quantum 
mechanics predicts that the result for the second measurement of $A$ must be 
random, while in our model, it will agree with the first result up to possible 
probabilistic state changes (i.e., in our model, the probability distribution 
from which the value of $A$ is drawn will not differ whether it is the third 
measurement in the sequence $AAA$ or in the sequence $ACA$).
However, Kochen-Specker tests are always carried out within compatible sets of 
observables, and, since we are (for the moment) only investigating what can be 
concluded from such a test alone, this is not our concern here.  Our main point 
is that this simple model can invalidate some
ideas to make Kochen-Specker tests robust against noise.  

Let us now consider what happens during a measurement of the left-hand side of 
Eq.~\eqref{CHSH} if violations of compatibility are present. Then, if we denote 
by $A_i$ the observed value of $A$, given that the hidden variable state is 
$\lambda_i$, $\ex{\chi_{\mathrm{CHSH}}}$ can be calculated as follows:
\begin{eqnarray}
 \nonumber\ex{\chi_{\mathrm{CHSH}}}&=&\sum_{i,j}(A_iB_j+B_iC_j+C_iD_j-D_iA_j)p_{ij}\\
 &\equiv&\sum_{i,j}K_{ij}p_{ij},
\end{eqnarray}
where the $p_{ij}$ denote the probability that the evolution of the system is $\lambda_i\to\lambda_j$, that is, that the state during the first 
measurement was $\lambda_i$, which transitioned to $\lambda_j$ before the 
second one, and we have introduced the quantity 
$K_{ij}=A_iB_j+B_iC_j+C_iD_j-D_iA_j$. The maximum $K^{\max}$ of the 
coefficients $K_{ij}$ provides the upper bound
\begin{equation}
 \ex{\chi_{\mathrm{CHSH}}}=\sum_{i,j}K_{ij}p_{ij}\leq\underset{ij}{\max}\lbrace 
K_{ij}\rbrace\equiv K^{\max}.
\end{equation}
Each $K_{ij}$ represents the value of $\chi_{\mathrm{CHSH}}$, given the hidden variable evolution $\lambda_i\to\lambda_j$. It is easy to check that 
$K_{0,8}=4$: $\lambda_0=(++\phantom{}\!++)$ and $\lambda_8=(-+\phantom{}\phantom{}\!++)$, and thus, $\langle AB \rangle = \langle BC \rangle = 
\langle CD \rangle = +1$, while $\langle DA\rangle = -1$. Hence, a simple model that after each measurement changes the system's state from 
$\lambda_0$ to $\lambda_8$ will maximally violate the CHSH inequality; if the change happens only with a certain probability $p$, obviously any 
value between $2$ and $4$ can be achieved.

It should be noted that despite the evolution of the 
hidden variable, this model is noncontextual in the sense that whether or not a state 
transition is effected does not depend on the measured context.  It thus seems 
surprising that this model can violate the CHSH inequality, apparently 
indicating contextual behavior. However, strictly speaking, noncontextuality 
simply does not apply in this case, as it is defined only under the assumption 
of perfect compatibility.

\section{Connection with previous works}
An approach to rein in the effects of compatibility violations was proposed in 
Ref.~\cite{errterms}. There, several classes of error terms were proposed, such 
that additional measurements may be performed in order to quantify the degree 
of failure of \emph{a priori} compatible observables to be compatible in the 
actual experiment, i.e., the degree of influence a measurement of $A$ has on 
the compatible measurement $B$, for example.
We will concentrate, for the moment, on the first class of error terms from Ref.~\cite{errterms}, which are those that have been experimentally 
implemented.

Based on an assumption of noise cumulation, that is, an assumption that 
additional measurements always lead to additional noise and thus a worse 
violation of compatibility, the inequality
\begin{eqnarray}\label{err}
 \nonumber\ex{\chi_{\mathrm{CHSH}}} - p^{\mathrm{err}}[BAB] - p^{\mathrm{err}}[CBC] \\- p^{\mathrm{err}}[DCD] - p^{\mathrm{err}}[ADA]\leq2,
\end{eqnarray}
holds \cite{errterms}. Here for instance $p^{\mathrm{err}}[BAB]$ is the probability that the second measurement of $B$ in the sequence $BAB$ 
disagrees with the first one.

However, it is clear that the model we discuss does not obey the assumption of 
cumulative noise: for an evolution such as $\lambda_0\to\lambda_4
\to\lambda_0$, clearly both measurements of $B$ in the sequence $BAB$ agree, 
but if $B$ were measured in the second place of the sequence, then it would 
have yielded a value opposite to the first. Thus, the model is not necessarily 
constrained by Inequality \eqref{err}; and in fact, since the error terms all 
vanish for such an evolution, it is clear that the model can violate it.

Alternatively, it may be noted that while the original CHSH-inequality is only concerned with measurement sequences of length 2, the error terms 
contain only sequences of length 3, and thus, can only provide information about the system's behavior during such sequences. This criticism holds
for the other two classes of error terms in Ref.~\cite{errterms} as well.

\section{Modified Inequalities} However, another approach, which does not need 
any additional measurements or further assumptions, is possible.  This amounts 
to essentially applying the ideas of Leggett and Garg to contextuality 
inequalities. Rather than employing the original inequalities proposed in 
Ref.~\cite{LG}, it is convenient for our purposes to use a slightly different 
formulation. Consider two different measurements $C$ and $C^{\prime}$, 
performed at two points in time. Then, 
$C(C+C^{\prime})+C^{\prime}(C-C^{\prime})=\pm2$, and thus
$
\ex{C^{\prime}C}+\ex{CC}+\ex{CC^{\prime}}-\ex{C^{\prime}C^{\prime}}\leq2,
$
where $\ex{CC^{\prime}}$ denotes the correlation between $C$, measured at 
$t_1$, and $C^{\prime}$, measured at $t_2$.

We can now impose a similar time-ordering of observables on Eq.~\eqref{CHSH}, 
to get
\begin{equation}\label{CHSH*}
 \ex{\chi_{\mathrm{CHSH}}}=\ex{AB}+\ex{CB}+\ex{CD}-\ex{AD}\leq2.
\end{equation}

It is not hard to see that for Eq.~\eqref{CHSH*}, $K_{ij}\leq2$ for all 
$(i,j)$: if the first three terms are equal to $+1$, the fourth is necessarily 
equal to $+1$, as well. Hence, our model cannot violate Inequality 
\eqref{CHSH*}, despite the violation of compatibility. Since in the case of a 
trivial evolution of the hidden variables, i.e. an evolution that leaves the state invariant, we recover the usual notion of 
(sequential) noncontextuality, an experimental test of Eq.~\eqref{CHSH*} 
constitutes a test of quantum contextuality robust against the compatibility 
loophole.

\begin{table}
\centering
\begin{tabular}{|ccc|}
\hline
$A=\sigma_z\otimes \mathds{1}$ & $B=\mathds{1}\otimes\sigma_z$ & $C=\sigma_z \otimes \sigma_z$\\
$a=\mathds{1}\otimes\sigma_x$ & $b=\sigma_x\otimes\mathds{1}$ & $c=\sigma_x\otimes\sigma_x$ \\ 
$\alpha=\sigma_z\otimes\sigma_x$ & $\beta=\sigma_x\otimes\sigma_z$ & $\gamma=\sigma_y\otimes\sigma_y$\\ \hline
\end{tabular}
\caption{The Peres-Mermin square, with the Pauli matrices $\sigma_i$, and the $2\times2$ identity matrix $\mathds{1}$. The observables in all rows and 
columns commute, and the product of all rows and the first two columns is equal to $\mathds{1}$, while for the last column, $Cc\gamma=-\mathds{1}$.}
\label{PM}
\end{table}

It should be noted that the CHSH-inequality is not the only one that can be modified to hold in the case of compatibility violations: another 
important inequality proposed in Ref.~\cite{Adan} is based on the Peres-Mermin square (\cite{PMsquare}; see Table~\ref{PM}).
Using the same reasoning as in the CHSH-case, the inequality
\begin{eqnarray}\label{PM*}
 \nonumber\ex{\chi_{\mathrm{PM}}}&=&\ex{ABC}+\ex{cab}+\ex{\beta\gamma\alpha}\\ &+& \ex{Aa\alpha}+\ex{\beta Bb}-\ex{c\gamma C}\leq4
\end{eqnarray}
holds also in the case of imperfect compatibility between observables. In quantum mechanics, a value of
$\langle\chi_{\mathrm{PM}}\rangle = 6$ can be reached. Again, it is here the ordering of the measurement sequences that matters: 
the original inequality proposed in Ref.~\cite{Adan} followed the ordering indicated in Table~\ref{PM}; but in this form, it is not hard to see 
that the inequality can be violated easily by our model. Interestingly, the ordering proposed here is also useful if the Mermin-Peres inequality 
should be used for estimating the dimension of a quantum system \cite{dimwit}.

The importance of this scenario is that this inequality is state-independent, that is, one
does not require a special quantum state for the violation (as is the case for 
the CHSH-inequality).  Furthermore, an experiment using sequential measurements 
on trapped ions already implemented this scenario by measuring the observables 
in Table~\ref{PM} in all possible permutations \cite{nature}.  This 
experiment focused on the violation of the inequality as originally proposed in 
Ref.~\cite{Adan}, and using this data, the observed value for Eq.~\eqref{PM*} is 
$\ex{\chi_{\mathrm{PM}}}=5.35(4)$.

Not every noncontextuality inequality can be modified this straightforwardly, 
though. For example, the Klyachko-Can-Binicio\u{g}lu-Shumovsky inequality 
\cite{Klyachko}
\begin{align}\label{KCBS}
 \nonumber\ex{\chi_{\mathrm{KCBS}}}=&\ex{AB}+\ex{BC}+\ex{CD}\\&+\ex{DE}+\ex{EA}\stackrel{NCHV}{\ge} -3 \stackrel{QM}{\ge} 5-4\sqrt5,
\end{align}
which exhibits a quantum violation even for a single qutrit system, as demonstrated experimentally in Ref.~\cite{Zei2}, cannot 
be rearranged appropriately. Nevertheless, our approach can be generalized: the modified inequality 
\begin{align}\label{KCBS*}
 \nonumber\ex{AB}&+\ex{CB}+\ex{CD}+\ex{ED}+\ex{EA}-\ex{AA}\\&\stackrel{NCHV}{\ge} -4 \stackrel{QM}{\ge} 4-4\sqrt5
\end{align}
holds for any noncontextual hidden-variable evolution. The reason for this is 
that it enforces the ordering conditions as in Eq.~\eqref{CHSH*}: to maximize 
the left hand side of Eq.~\eqref{KCBS*}, for instance $A_i$ must equal $E_j$, 
as must $A_j$; however, then $A_i=A_j$, and thus, $\ex{AA}=1$. This shows that 
even in the case of a single qutrit a Kochen-Specker test ruling out our model 
can be undertaken. However, one should note that due to this modification,
the relative quantum violation shrinks, since the absolute violation stays the same, 
while the absolute value of the classical expectation increases. Finally, it should be noted that a similar inequality like
Eq.~\eqref{KCBS*} has already been used in Ref.~\cite{Zei2} in order
to compensate for the fact that in this setup the observable $A$ was 
implemented in two different ways.

In fact, a recently proposed state-independent inequality violated by a single qutrit system \cite{YO} can be treated in the same way. This 
inequality features 13 observables $\lbrace A^1,\ldots, A^{13} \rbrace$, and the form that yields the maximum quantum violation is \cite{YOopt, YOexp}
\begin{equation}
 \sum_i\Gamma_i\ex{A^i}+\sum_{ij}\Gamma_{ij}\ex{A^iA^j}\leq16,
\end{equation}
where the coefficients are as follows: $\Gamma_i=1$ for $i\in\lbrace 4,7,10,\ldots,13\rbrace$, $\Gamma_i=2$ for $i\in\lbrace 1,5,6,8,
9\rbrace$, and $\Gamma_i=3$ for $i\in\lbrace 2,3\rbrace$; $\Gamma_{ij}=-1$ for 
$(i,j)\in\lbrace (1,2)$, $(1,3)$, $(1,4)$,
$(1,7)$, $(4,10)$, $(8,10)$, $(9,10)$, $(5,11)$, $(7,11)$, $(9,11)$, $(6,12)$, $(7,12)$, $(8,12)$, $(4,13)$, $(5,13)$, $(6,13)\rbrace$, 
$\Gamma_{ij}=-2$ for $(i,j)\in\lbrace 
(2,3),\linebreak[1](2,5),\linebreak[1](2,8),\linebreak[1](3,6),
\linebreak[1](3,9),\linebreak[1](5,8),\linebreak[1](6,9)\rbrace$, and 
$\Gamma_{ij}=0$ else.
By checking all possible hidden variable evolutions one verifies that the modified inequality \footnote{There is no ordering ambiguity for
the single measurements $\langle A^i\rangle$ since we consider disturbances to result from measurement interactions; a single measurement
can always be considered to occur in the first position of a measurement sequence.}
\begin{equation}
 \sum_i\Gamma_i\ex{A^i}+\sum_{ij}\Gamma_{ij}\ex{A^iA^j}+4\sum_i\ex{A^iA^i}\leq68
\end{equation}
cannot be violated by noncontextually evolving models.
However, since the maximum quantum value in 
this case is only $69+\frac{1}{3}$, the relative violation is reduced to 
$\frac{1}{51}\approx1.96\%$, compared to originally $\frac{1}{12}\approx8.3\%$.

\section{Conclusion} 
We have provided a novel approach to the compatibility problem 
in Kochen-Specker experiments. Using the idea of time-ordering,
as first proposed by Leggett and Garg, we have derived new 
inequalities violated by quantum mechanics even in the case 
of imperfectly compatible measurements. This shows that with
a careful ordering of the measurements classical models
can be ruled out, which cannot be excluded with existing approaches
\cite{errterms}. Nevertheless, we are not claiming that our modified inequalities allow a test of the Kochen-Specker theorem free from
the compatibility loophole. Our results, however, show that with a simple reordering
of the measurements a significantly larger class of hidden-variable models can 
be ruled out.

We thank C. Roos for providing us with the data from their 
experiment, and C. Budroni and D. Bruß for valuable discussions. 
This work has been supported by the EU (Marie Curie CIG
293993/ENFOQI) and BMBF (projects QuOReP and QUASAR).

\end{document}